\documentclass[preprint,authoryear,12pt]{elsarticle}

\journal{Astronomy and Computing}

\usepackage[]{algorithm}
\usepackage{algpseudocode}
\usepackage{amssymb}

\usepackage{graphicx}
\usepackage{amsmath}
\usepackage{longtable}
\usepackage{color}
\usepackage{float}

\newcommand{\be}{\begin{equation}}
\newcommand{\ee}{\end{equation}}
\newcommand{\bea}{\begin{eqnarray}}
\newcommand{\eea}{\end{eqnarray}}
\newcommand{\etal}{et al.}

\newcommand{\balpha}{\mbox{\boldmath$\alpha$}}
\newcommand{\bbeta}{\mbox{\boldmath$\beta$}}
\newcommand{\btheta}{\mbox{\boldmath$\theta$}}

\begin{document}

\begin{frontmatter}

\title{ Accelerating gravitational microlensing simulations using the Xeon Phi coprocessor}



\author[a]{Bin Chen\corref{author}}
\ead{bchen3@fsu.edu}
\cortext[author] {Corresponding author. 
}

\author[b]{Ronald Kantowski}
\ead{kantowski@ou.edu}
\author[b]{Xinyu Dai}
\ead{xdai@ou.edu}
\author[b]{Eddie Baron}
\ead{baron@ou.edu}
\author[a]{Paul Van der Mark}
\ead{pvandermark@fsu.edu}

\address[a]{Research Computing Center, Florida State University, Tallahassee, FL 32306, USA}
\address[b]{Homer L. Dodge Department of Physics and Astronomy, The University of Oklahoma, Norman, OK, 73019, USA}

\begin{abstract}
Recently Graphics Processing Units (GPUs) have been used to speed up very CPU-intensive  gravitational microlensing simulations.
In this work, we use the Xeon Phi coprocessor to accelerate such simulations and compare its performance on a microlensing code with that of NVIDIA's GPUs.
For the selected set of parameters evaluated in our experiment, we find that the speedup by Intel's Knights Corner coprocessor is comparable to that by NVIDIA's Fermi family of GPUs with compute capability 2.0,
but less significant than GPUs with higher compute capabilities such as the Kepler.
However, the very recently released second generation Xeon Phi, Knights Landing, is about 5.8 times faster than the Knights Corner, and about 2.9 times faster than the Kepler GPU used in our simulations.
We conclude that the Xeon Phi is a very promising alternative to GPUs for modern high performance microlensing simulations.
\end{abstract}



\end{frontmatter}


\section{Introduction}

Gravitational microlensing was first discussed by \cite{Refsdal64} and discovered in the quadrupole quasar lens system Q~2237+0305 by \cite{Irwin89}.
It has since become a powerful probe constraining the properties of both background sources and foreground lenses.
For example, microlensing observations combined with Bayesian Monte Carlo simulations \citep{csk04} have been used to constrain the emission sizes of the quasar accretion disk at various wavelengths \citep{Pooley06,Chartas09,Dai10,Morgan12,Ana13,Blackburne14,Blackburne15,MacLeod15,Guerras16}.
It can also be used to constrain foreground lens properties such as stellar/dark matter fractions \citep{Mao98,Mediavilla09}.
The large synoptic survey telescope (LSST) is expected to discover thousands of new strong lens systems in the next decade, and microlensing should exist in many of these new lens systems \citep{Abell09}.
A joint analysis of many microlensing systems requires the exploration of a large parameter space but could produce useful constraints on the cosmological parameters \citep{Thompson09,Bate10,Fluke13,Fluke14}.
Such a large campaign poses new challenges not only to observers, but also to computational astrophysicists and data scientists \citep{Vohl15}.

When light from a background source (e.g., a quasar) passes a foreground lens galaxy or cluster of galaxies, it is gravitationally lensed by both the galaxy (or cluster) as a whole, and by the many stars around the light beam.
The gravitational bending angle by a stellar mass lens is of the order of a micro arcsecond, and is what one calls microlensing.
The gravitational microlensing equation of a random star field can be written as \citep{Schneider92}
\be\label{lens_eq}
\mbox{\boldmath$\beta$} =
\begin{pmatrix}
    1-\kappa_c-\gamma_1& -\gamma_2    \\
    -\gamma_2                  & 1 -\kappa_c +\gamma_1
\end{pmatrix} \mbox{\boldmath$\theta$}
-\sum_{i=1}^{N_*}\frac{m_i(\mbox{\boldmath$\theta$}-\mbox{\boldmath$\theta$}_i)}{(\mbox{\boldmath$\theta$}-\mbox{\boldmath$\theta$}_i)^2}\,,
\ee or as
\be
\mbox{\boldmath$\beta$} = \mbox{\boldmath$\theta$} - \mbox{\boldmath$\alpha$},
\ee where
\be\label{bending}
 \mbox{\boldmath$\alpha$}=
 \begin{pmatrix}
 \kappa_c+\gamma_1 & \gamma_2 \cr
  \gamma_2                 & \kappa_c-\gamma_1
\end{pmatrix}
\mbox{\boldmath$\theta$}
 +\sum_{i=1}^{N_*}\frac{m_i(\mbox{\boldmath$\theta$}-\mbox{\boldmath$\theta$}_i)}{(\mbox{\boldmath$\theta$}-\mbox{\boldmath$\theta$}_i)^2},
 \ee is the gravitational bending angle, $\mbox{\boldmath$\beta$}$ and \mbox{\boldmath$\theta$} are respectively the source and image angular positions. All three angles are in units of the Einstein ring angle $\theta_E$ of a unit solar mass lens
\be
\theta_E\equiv \sqrt{\frac{D_{\rm ds}}{D_{\rm d}D_{\rm s}}\frac{4GM_{\odot}}{c^2}},
\ee  $\kappa_c$ is the smooth surface mass density, $\gamma=(\gamma_1,\gamma_2)$ is the external shear, and $m_i$ is the mass of the $i^{\rm \,th}$ micro-lens in units of the solar mass $M_\odot$.
In the above equation $D_{\rm ds},$ $D_{\rm d}$ and $D_{\rm s}$ are the angular diameter distances from the lens to the source, from the observer to the lens and to the source, respectively.
The microlensing equation of a random star field can not be analytically solved, and numerical ray-tracing is the standard method to generate simulated magnification patterns in the source plane \citep{Schneider87,Wambsganss99}.
In practice, a large number of rays are traced backward from the observer to the lens plane, and then to the source plane (therefore the name ``backward raytracing").
The number of rays collected in a given pixel in the source plane is proportional to the gravitational lensing magnification/demagnification for a source at that pixel's location.
Given a model for the background source size and intensity profile, the magnification pattern can be used to generate simulated microlensing light curves, which when compared with observational data can be used to constrain the properties of both the foreground lens and the background source.
For example, the quasar X-ray emission size has been recently constrained to be about $10\,r_g$ (gravitational radius) of the central supermassive black hole using \emph{Chandra} data \citep{Chen11,Chen12}  and the Bayesian Monte Carlo analysis \citep{csk04,Chartas09,Dai10,Morgan12,Ana13,Blackburne14,Blackburne15,MacLeod15,Guerras16}.
 An important step of the Bayesian Monte Carlo microlensing analysis is to generate a large simulated magnification map ($\sim$$20\theta_E\times 20\theta_E$) of high resolution ($\sim$$10^4\times 10^4$ pixels).
 To constrain the multi-dimensional parameter space of a foreground lens a large number of such maps needs to be generated.

The Central Processing Unit (CPU)-intensive part of microlensing ray-tracing is clearly the summation over the multitude of stars in Equation~(\ref{lens_eq}). This summation has to be repeated for many rays, i.e., for a large number of image angles $\btheta_{i,j}$  covering a 2D grid at the image plane.
The number of stars $N_*$ in each summation is proportional to the stellar surface mass density $\kappa_*$ and the area of the lens window.
The total number of rays traced is proportional to the area of the lens window (equivalently, the total number of pixels in the lens window) and the number of rays chosen for each pixel (the ray density).
A ray density of a few hundred to a couple of thousand is typically required to reduce the statistical error.
Fixing the ray density while increasing the lens window area by a factor of 2 will increase $N_*$ by a factor of 2 and the number of rays by a factor of 2, thus increasing the computing time by a factor of 4.
Fixing the ray density and lens window size, while increasing $\kappa_*$ by a factor of 2 will double the number of stars $N_*$ and thus double the computing time.

To compute many bending angles ($\sim$$10^{10}$) by many stars (from a few hundred to many millions) by brute force on a serial computer is very time-consuming.
To speed up  microlensing simulations, either smart algorithms which avoid brute force summations [see e.g., \citet{Mediavilla11}] or parallel computing is called for.
One important method to avoid the brute force summation is the so-called {\it hierarchical tree algorithm} \citep{Wambsganss99,Barnes86}.
In the current paper, we focus on the other option, i.e., we speed up the gravitational microlensing simulation via parallel computing,
in particular, parallel computing using hardware accelerators/coprocessors: Graphics Processing Units (GPUs) and the Xeon Phi.
GPU microlensing raytracing has been studied very recently \citep{Thompson09,Bate10,Fluke13,Fluke14}.
Here we evaluate the performance of parallel gravitational microlensing ray-tracing using the Xeon Phi and compare it with the GPU version of the same algorithm.

The plan of the paper is as the following:
In Section~\ref{sec:intro} we introduce the Xeon Phi coprocessor.
In Section~\ref{sec:algo} we outline the parallel ray-tracing algorithm for the microlensing simulation used in this work.
In Section~\ref{sec:experiment} we define the experiment we conduct to compare the performance of the Xeon Phi coprocessors with the NVIDIA GPUs.
In Section~\ref{sec:result} we summarize our results, and 
in Section~\ref{sec:discuss} we discuss 
and conclude.

\section{Parallel microlensing simulation via the Xeon Phi}\label{sec:Phi}

\subsection{An introduction to the Xeon Phi}\label{sec:intro}

The Xeon Phi is a family of many-core parallel coprocessors manufactured by Intel for high performance computing.
The first release of the Xeon Phi was the Knights Corner (KNC) which contains some 57 to 61 cores built on the Intel Many Integrated Core (MIC) micro-architecture.
A MIC coprocessor core is more complicated/expensive than a GPU core, but it is much simpler than a single core of a Xeon processor.
The KNC coprocessor is used as an accelerating device for a traditional computer node, similar to the way a GPU card is used as an accelerator for a computer node.
A KNC Phi card can have 6/8/16 Giga-Bytes (GB) of high speed memory ($\sim$320 GB/s) which communicates with a host node through the Peripheral Component Interconnect Express (PCIe) bus (16 GB/s). 
The most recent release of Xeon Phi, Knights Landing (KNL), can run as an independent computer node with a self-booting Linux operating system (i.e., not attached to a Xeon host node anymore).
Xeon Phi scales up parallel applications through its many cores per processor, multi-threading per core (optimal performance at $\sim$4 threads per core), and its powerful vector processing units  (e.g., 512-bit vector registers). 
A big advantage of Xeon Phi is that it supports common programming languages such as Fortran, C/C++, and parallel programming models such as the distributed memory programming model Message Passing Interface (MPI), the shared memory programming model Open Multi-Processing (OpenMP), and the hybrid programming model  MPI+OpenMP. 
Applications can run in either the native or the offload mode.
In the native mode, the application is compiled on the host, but is launched and run directly from the Xeon Phi coprocessor.
In the offload mode, the application is compiled and launched from the host, but the CPU-intensive part is offloaded to the Phi coprocessor during the runtime \citep{Jeffers13}.
We focus on the offload mode in the current work, since this programming model is very similar to the GPU programming model where the application is also launched from the host but with the CPU-intensive part accelerated by the GPU.

Table~\ref{tab:hardware} compares the hardware specifics of a multi-core Sandy Bridge (SB) host node (Intel Xeon E5--2670 CPU; Dell C8220 motherboard), a Xeon Phi coprocessor (KNC), and an NVIDIA GPU (Tesla M2050) used in this work (more information can be easily found from the vendor's website using the product name).
The Xeon Phi coprocessor used in this work is a special version of KNC delivered to the Texas Advanced Computing Center. 
It is similar to the Xeon Phi 5120D, but with 61 cores (instead of 60) and a slightly higher clock frequency.\footnote{Refer to https://portal.tacc.utexas.edu/user-guides/stampede for more detail.} 
The SB host node has 16 cores, much fewer than the Xeon Phi (61 cores) and Tesla M2050 (448 cores).
However, its memory (32 GB) is much larger than the Xeon Phi (8 GB) and Tesla M2050 (3 GB).
The Xeon Phi has the highest peak memory bandwidth, 320 GB/s, compared with the SB (102 GB/s) or the Tesla M2050 (148 GB/s).
The theoretical peak double precision performances of the host node, Xeon Phi, and GPU are 0.333, 1.065, and 0.515  Tera Floating-point Operations Per Second (TFLOPS), respectively.
The Tesla M2050 GPU has a Fermi architecture with compute capability 2.0.

\begin{table}[t]
\scriptsize
\caption{
\label{tab:hardware}
Host Computer node versus Xeon Phi and Gpu.
}
\begin{tabular}{cccc}
\hline
	& Sandy Bridge & Xeon Phi  & Gpu   \\
	& E5--2670 & KNC\textsuperscript{*} & Tesla M2050 \\
\hline
Clock Frequency (GHz)   	 	&   2.60    			& 1.10  				& 1.15    	\\
Number of Cores         	 	&   16    			& 61  				& 448    	\\
Memory Size/Type       	 	&   32GB/DDR3    	& 8GB/GDDR5  		& 3GB/GDDR5    \\
Memory Clock (GHz)       		&   1.6    			& 1.375  				&  1.55   	\\
Peak DP (TFLOP/s)	 		&   0.333    		& 1.065  				&  0.515     \\
Peak Memory Bandwidth (GB/s)&   102    			& 320   				&  148    		\\
Host-Coprocessor Interconnect &    $\cdots$		& PCIe2.0x16 (16GB/s)	& PCIe2.0x16 (16GB/s)	\\
\hline
\multicolumn{4}{l}{\textsuperscript{*}\scriptsize{The KNC used is a special version similar to Xeon Phi 5120D, but with 61 cores. }}
\end{tabular}
\end{table}

\subsection{Parallel ray-tracing Algorithm}\label{sec:algo}

An outline of the simple (brute force) parallel microlensing simulation is shown in {\bf Algorithm 1.}
The ray-tracing code is parallelized using the shared memory programming model OpenMP.
The host computer (a SB node with two 8-core processors) and the Xeon Phi coprocessor each has their own distributed memory (Table~\ref{tab:hardware}).
The 32 GB memory on the host is shared among its 16 cores, and the 8 GB memory on the coprocessor KNC is shared among its 61 MIC cores,
but there is no memory sharing between the host and Phi cores.
Consequently, the OpenMP code can run either solely on the host, or be offloaded to the coprocessor.
To use both the host and coprocessor for the parallel ray-tracing, a distributed memory programming model such as MPI is required.
For example, a few MPI tasks can be launched on both the host and the Phi card, each of these MPI tasks can spawn a few OpenMP threads to do part of the ray-tracing.
In the end, the results from each MPI task are collected and reduced to obtain the final magnification map.
For situations where many independent simulations need to be generated, for example, simulations with a grid of input parameters such as $\langle M_*\rangle,$ $\gamma,$ $\kappa_*$, $\kappa_c,$ etc., heterogeneous computing using both the host and Phi card can be easily realized by letting the host and Phi each work on their own simulations.
In this paper, we present the simple case where we generate only one simulation map using either the host or the coprocessor in the offload mode.
This makes the performances of the host node, the Xeon Phi coprocessor, and the NVIDIA GPU accelerator easily comparable.
To design a hybrid code (MPI+OpenMP) with load-balancing mechanism should be straightforward once the performance of the host and the Phi node are well understood.
To compile the code for running on the host (instead of the coprocessor), simply comment out the line with the {\it \#pragma offload} compiler directive from {\bf Algorithm 1}.

\begin{algorithm}[H]
\caption{Pseudo-algorithm for microlensing ray-tracing.
Pragmas for OpenMP and Xeon Phi offloading is colored.  }
\begin{algorithmic}
\State{\color{blue}\#pragma offload target(mic)}  \Comment{Offload to Xeon Phi Coprocessor}
\State{\color{blue}\#pragma omp parallel for collapse(2)} \Comment{OpenMP parallelization}
\For{ $j = 0: \rm n\_rays\_y -1$}
\For{ $i = 0: \rm n\_rays\_x -1$}
  \State{evaluate the image angle $\btheta_{i,j}$}
  \State{$\balpha_{i, j} = \balpha_0(\btheta_{i,j};\kappa,\gamma)$} \Comment{The first term in Equation~(\ref{bending})}
   \State{\color{blue}\#pragma simd} \Comment{SIMD vectorization for Xeon Phi}
   \For{ $k = 0: N_* - 1$}
  \State{ $\balpha_{i,j}$ += $\balpha_{k}(\mbox{\boldmath$\theta$}_{i,j},\mbox{\boldmath$\theta$}_k)$}  \Comment{$\alpha_{k}$ the bending angle by $k$-th star}
  \EndFor
  \State{$\bbeta_{i,j} =\btheta_{i,j} -\balpha_{i,j}$} \Comment{$\bbeta_{i,j}$ the source angle}
  \State{convert $\bbeta_{i,j}$ into pixel coordinates $(i_s, j_s)$}
  \If{ $(i_s, j_s)$ in the source window }
\State{\color{blue}\#pragma omp atomic}
\State{ $\rm mag[j_s, i_s]$ += 1} \Comment{$\rm mag[\cdot,\cdot]$ the 2D magnification pattern}
\EndIf
\EndFor
\EndFor
\end{algorithmic}
\end{algorithm}


In order to compare the performance of Intel's Xeon Phi with NVIDIA's GPU accelerator, we created a GPU version of the microlesing code using the programming language CUDA.
The algorithm is essentially the same as {\bf Algorithm 1,} except that the two outer for loops are now peeled off from the code,
a CUDA kernel function is created for the inner-most for loop, the kernel is invoked with a 2D grid of blocks, each block with a 2D thread structure [for basics of the 
CUDA programming model, please refer to \cite{Cheng14}].
For the current work, programing the CUDA GPU code is slightly more complicated than programming the Phi code, mainly because CUDA requires the user to explicitly handle the data transfer between the host and the accelerator.
For example, the input data for the microlensing ray-tracing (such as the data structure storing the random star field) is generated on the host, and must be explicitly allocated and copied from the host to the accelerator. After the computation on the GPU accelerator is finished, the magnification map data must be copied from the accelerator back to the host, and then deallocated from the accelerator's memory space.
Since the Xeon Phi supports a general programming model with multiple programming languages (Fortran, C/C++, etc.), the majority of the offloading and data management is done automatically (implicitly) by the compiler.
For example, specifying the {\it \#pragma offload out(mag:length(size))} directive will inform the compiler to create code copying the magnification map data array ``mag[size]" back to the host after the Phi coprocessor has finished the computation.

As a summary, both the Xeon Phi (KNC) and the GPU version of the microlensing simulation run in the ``offload" mode, i.e., the input data (e.g., the random star field) is generated on the host, then copied to the coprocessor/accelerator implicitly/explicitly.
All computations, i.e., Equation~(\ref{bending}), are done on the coprocessor or accelerator, and in the end the data is copied back to the host. 
To run the microlensing simulation in the ``offload" mode is justified by the fact that the amount of data movement between the host and coprocessor/accelerator is minimal, and the computing time is much longer than the communication overhead.
For example, a large integer magnification matrix of dimension $10^4\times 10^4$ is only 0.4 GB, copying this data to the host through PCIe 2.0 bus takes less than 1 second, and this data movement needs to be done only once.
This overhead is negligible compared to the microlensing simulation, which can take many hours.

\subsection{Experiment of performance comparison}\label{sec:experiment}

Our experiment is designed as the following: 
First, we created two separate programs using the {\bf Algorithm 1}: an OpenMP code written in C which can run either solely on the host node, or run on the Xeon Phi under the offload mode, and a CUDA version for the NVIDIA GPUs. 
We run the OpenMP and CUDA codes on the host, host+Xeon Phi, and host+GPU,  to generate a large microlensing magnification map using exactly the same set of input parameters.
The three types of runs produce the same magnification map (i.e., the integer image matrices counting the number of rays collected for each pixel are exactly the same).
We next compare the multi-threading performance of the host node and the Xeon Phi with and without vector support.
We use the run time of the single-thread (serial) host computer code without vector support as the baseline.
The speedup obtained by turning on multi-threading and/or vector support is defined to be the quotient of the longer baseline run time by the shorter parallel run time.  
We next compare the performances across the host node, Xeon Phi, and the GPU.
For this comparison we use the best performance of each code. 
For example, the Xeon Phi code performs the best when the vector support is turned  on and when we overload each MIC core with multiple threads.
We use the best case for the SB host node (16 threads with vector support) as the baseline. 
The speedup by Xeon Phi or GPU is defined in a similar way. 
We compare the performance of Xeon Phi with NVDIA GPU using two generations of each product, i.e., KNC (2013) and KNL (2016) for the Xeon Phi, Fermi (compute capability 2.0) and Kepler (compute capability 3.5) for the NVIDIA GPUs.

\section{Results}\label{sec:result}

\begin{figure*}
\begin{center}
\includegraphics[width=1.0\textwidth,height=0.5\textheight]{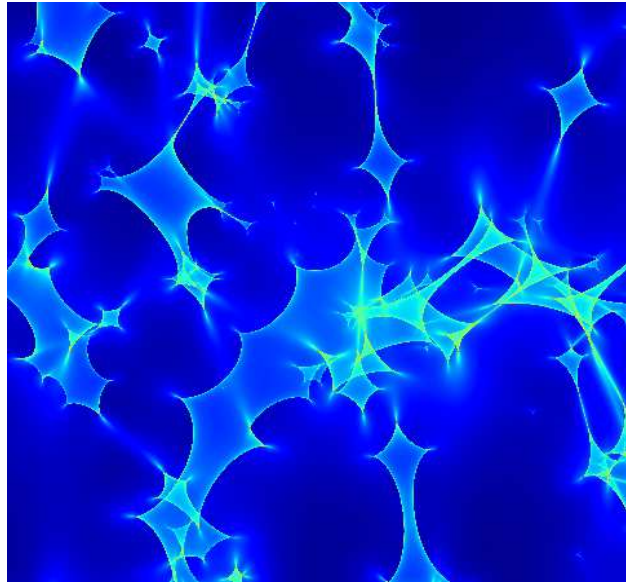}
\end{center}
\caption{A gravitational microlensing magnification map generated by the parallel ray-tracing code. 
We have assumed a random star field with mean lens mass $\langle M_*\rangle=0.3 M_\odot$, surface mass densities $\kappa_{c}=0.4,$ $\kappa_*=0.2,$ and external shear $\gamma=(0.2,0).$
The image is of high resolution $6400\times 6400$ and size about $12\theta_E\times 12\theta_E$ where $\theta_E$ is the Einstein ring angle of a lens of mass $\langle M_*\rangle$.
The host computer node, Xeon Phi coprocessor, and GPU accelerator produce exactly the same magnification map.
}
\label{fig:shear}
\end{figure*}

Figure~\ref{fig:shear} shows a simulated microlensing magnification pattern by a random star field with mean lens mass $\langle M_*\rangle=0.3 M_\odot,$
surface mass densities $\kappa_{\rm c}=0.4$ (smooth mass), $\kappa_*=0.2,$ and external shear $\gamma=(0.2,0).$ 
For simplicity, we have assumed a log-uniform lens mass distribution with $0.01M_\odot<M<1.6M_\odot$.
The image size is about $12\theta_E\times 12 \theta_E$ where $\theta_E$ is the Einstein ring angle of the mean lens mass $\langle M_*\rangle.$
Such a choice of parameters is typical for probing the AGN X-ray emission size using gravitational microlensing techniques \citep{Chen13a,Chen15}.
The diamond-like curves are caustics where the lensing magnification diverges for point sources.
The host computer node, Xeon Phi coprocessor, and GPU accelerator produce exactly the same magnification map.

\begin{figure*}
\begin{center}
\includegraphics[width=0.8\textwidth,height=0.5\textheight]{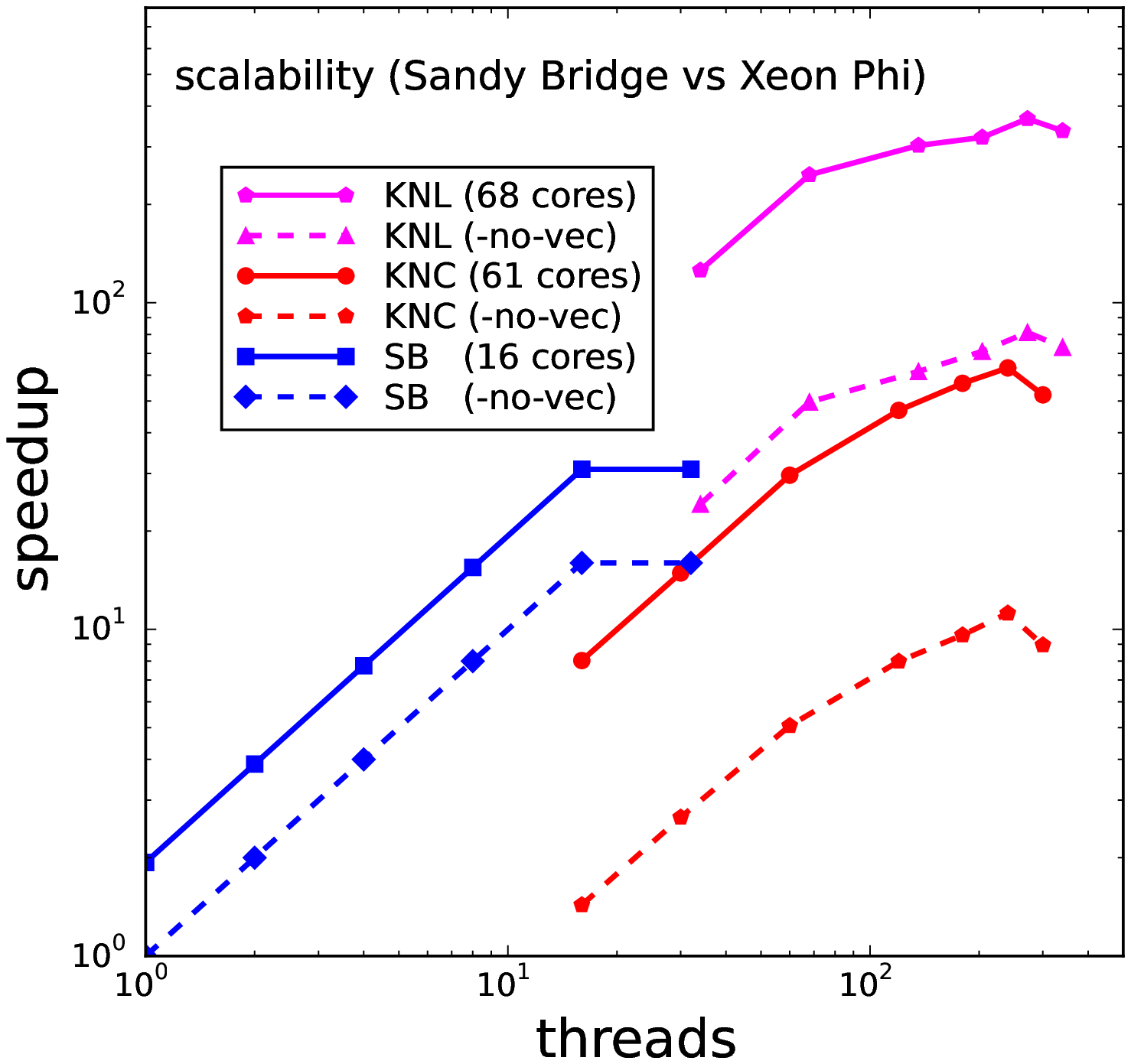}
\end{center}
\caption{Scalability of the Xeon Phi coprocessor.
The coprocessor KNL has 68 cores,  KNC has 61 cores, and the host SB node has 16 cores.
The magenta, red, and blue curves show the speedup via multi-threading for the KNL, KNC, and SB, respectively.
The solid and dashed curves show the results with and without vector support, respectively.
There is no performance gain when overloading each host core with 2 threads.
For the coprocessor, the maximum performance is achieved when each core runs 4 threads.
Turning on the vector support for the host produces a $\sim$$1.9$ speed up, whereas a vectorization speedup of $\sim$$5.8$ and $\sim$$4.5$ is achieved for the KNC and KNL, respectively.
The speedup of KNC (61 cores, 240 threads) and KNL (68 cores, 272 threads) with respect to the SB host (16 cores, 16 threads) is about $2.04$ and $11.84,$ respectively (all with vector support turned on).
}
\label{fig:phi_scale}
\end{figure*}

The parallel speedup via OpenMP multi-threading on the SB host node and on the Xeon Phi coprocessor KNC is shown in Figure~\ref{fig:phi_scale}.
We also tested the performance of the microlensing code on the latest release (July 2016) of Xeon Phi, KNL. 
In Figure~\ref{fig:phi_scale}, the blue, red, and magenta curves show the speedup via multi-threading for the SB, KNC, and KNL, respectively.
The solid lines show the results with Single Instruction Multiple Data (SIMD) vectorization turned on, whereas the dashed curves show the performance without vector support  (see {\bf Algorithm 1}).
Turning on the vector support for the host produces a $\sim$$1.9$ speedup, whereas a vectorization speedup of $\sim$$5.8$ and $\sim$$4.5$ is achieved for KNC and KNL, respectively.
Being able to vectorize the code is more critical for the coprocessor than for the processor, given that the vector registers on the Phi are 512 bits wide (8 double precision floating point numbers), whereas the host processors have only 256 bit vector registers.
However, note that the 61 MIC cores in the coprocessor KNC have much simpler architecture than the host cores. 
For example, the coprocessor cores have in-order architecture, whereas the host cores have out-of-order architecture \citep{Patterson11}, and are running with a lower clock rate (1.1 GHz vs 2.6 GHz, see Table~\ref{tab:hardware}).
To hide the instruction latency for this simple (but still pipelined) MIC architecture, multi-threading is explicitly required for the Phi coprocessor (if each core runs only one thread, then half of the instruction cycles will be wasted because instructions from the same thread can not be scheduled back to back in the pipeline).
As can be seen from Figure~\ref{fig:phi_scale}, the best Phi performance is achieved when each core has 4 threads.
On the other hand, overloading the 16 host cores with 32 threads does not produce any performance gain.
As a whole, the speedup of the 61-core KNC with respect to a 16-core SB node is about $2.04$ (the 240 threads Phi vs the 16 threads host; both with vector support turned on).
A speedup of order $\sim$$2$ is quite normal when comparing the KNC and SB peak performance \citep{Jeffers13}.
The speedup by the KNL is about 11.84 with respect to the host processor.

\begin{figure*}[h]
\begin{center}
\includegraphics[width=0.8\textwidth,height=0.4\textheight]{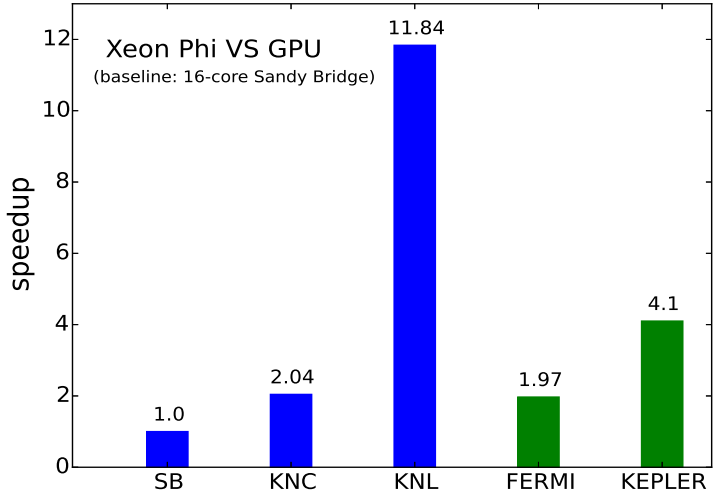}
\end{center}
\caption{ Performance of Xeon Phi versus GPU.
The best performance of the SB host node is used as the baseline.
The KNL (68 cores) is $\sim$5.8 times faster than the KNC (61 cores), and $\sim$2.9 times faster than Kepler GPU (K20) for the microlensing simulation.
}
\label{fig:phi_gpu}
\end{figure*}

Having scaled (through multi-threading) and vectorized the microlensing code for the Xeon Phi, we can now compare the performance of the Intel Xeon Phi and the NVIDIA GPU.
NVIDIA has released several generations of the GPUs, i.e., from Tesla to Fermi, Kepler, Maxwell, and Pascal.
We first compare the performance of the KNC (the first release of Xeon Phi by Intel in 2013) with the Tesla M2050 (an early version of NVIDIA GPU released in 2010 with the Fermi architecture), and find that
 the performance of KNC is about the same as Tesla M2050 (compute capability 2.0) for the microlensing simulation.
 For an example, the same simulation configuration for Figure~\ref{fig:shear} takes the KNC coprocessor 7,361 seconds, the Tesla 7,640 seconds, and the SB host 15,043 seconds.
However, running the same CUDA code on the Kepler GPU (K20, with compute capability 3.5) takes 3,669 seconds, which is twice as fast as KNC.
The performance of the KNL is about 5.8 times better than the KNC, and about 2.9 times better than the Kepler GPU used in this work, see Figure~\ref{fig:phi_gpu}.

\section{Discussion}\label{sec:discuss}

In this paper we have shown that gravitational microlensing simulations can be significantly accelerated using the Xeon Phi coprocessor (Figure~\ref{fig:phi_scale}).
We compared the performance of the Xeon Phi coprocessor with the NVIDIA GPU (Figure~\ref{fig:phi_gpu}).
We have focused on the KNC coprocessor, instead of the very recently released KNL, given that KNC is currently more commonly available than the KNL, and that KNC as a coprocessor very much resembles GPU as an accelerator.
Realizing essentially the same algorithm on Xeon Phi and GPU makes the comparison between the two hardware accelerators a fair one.
Programming for KNL is easier than for KNC, because KNL can run parallel code as a host (not as a coprocessor) and there is no need to offload code/data from a host (e.g., a SB node) to the Phi (co)processor.
Many legacy code need only be recompiled for the MIC architecture to run on the KNL (and future generations of Xeon Phi), and there is no need to rewrite the source code.
For the example presented in this work we have found that the performance of the KNC is comparable to that of the Fermi family of GPUs, but not as good as modern GPUs such as the Kepler.
However, the second generation Xeon Phi, KNL, is about 3 times faster than the Kepler GPU (K20, compute capability 3.5) used in this work.
Better performance is expected for future releases of the Xeon Phi.
We did not compare the performance of the Xeon Phi with GPUs more modern than the Kepler family, such as the Maxwell (compute capability 5.x), and the Pascal (compute capability 6.x). 
Given the significant performance gain when shifting from Fermi to Kepler (see Figure~\ref{fig:phi_gpu}), we expect the performance of the Maxwell and Pascal will be better than the Kepler. 
In particular, it would be interesting to see how the recently invented technologies, such as the NVLink high speed communications protocol, can boost the performance of  microlensing simulations.
 
The performance gain obtained from the Xeon Phi coprocessor (e.g., the KNC is about 2 times faster than a 16 core SB node) is typical according to many benchmark examples provided by the Intel \citep{Jeffers13}.
But the relative performance between the Xeon Phi and the GPU should depend on both the algorithms used and the specific region of the parameter space explored.
The dependence of a GPU microlensing code's performance on the parameter space has been investigated previously \citep{Thompson09, Bate10}. 
For example, \cite{Thompson09} first studied the GPU speedup as a function of the number of stars (from $\sim$10 to $10^8$), and \cite{Bate10} further investigated the GPU performance variation with respect to the external shear and the smooth matter fraction, etc.
In the current work we have chosen a relatively simple algorithm and parallelizing strategy which can be easily adapted to a multi-core CPU, a Xeon Phi coprocessor, and a GPU. 
We have run the simulation assuming a set of parameters which is typical for probing the AGN emission sizes in different wavelengths, in particular, the high energy X-ray band, using the microlensing techniques \citep{Chen13a,Chen15}.   
This does not mean that the algorithm used in this work is necessarily optimal for either Xeon Phi or GPU, or the numbers reported in this work comparing Xeon Phi with GPU (e.g., the KNL is about 3 times faster than K20) will remain valid for the whole parameter space.
Even for a simple brute force microlensing code, different parallelizing strategy can be applied depending on the hardware used and the region of the parameter space explored.
For example, in the current work the bending angle at one image position by all microlenses are computed by a single thread for both the Xeon Phi and the GPU (i.e., a CUDA device function for the GPU).
This is reasonable for the case presented in this paper where we need very high resolution magnification maps to resolve compact AGN X-ray coronae (a few $r_g$ of the central supermassive black hole) which requires many rays to be traced, but the number of stars needed for the simulation is only modest, e.g, from a few hundred to $\sim$$10^4$. 
But for cases with a very large number of microlenses, say, of order $10^9,$ as is needed for some cosmological microlensing simulations, packaging the computation of the bending angle  by billions of microlenses into a single thread is probably not a good choice for GPUs, given that a modern GPU like K20 can spawn millions of threads, in constant to Xeon Phi which runs with only a few hundred threads. 
For such cases, more flexible parallelizing strategy should be used, e.g., the bending angle by billions of stars could be split between multiple threads, the star field data could be loaded from the host memory to device memory in multiple batches, or even be split among multiple GPUs. 
A full exploration of such possibilities is beyond the scope of this work, and interested readers should refer to \citet{Thompson09} and \citet{Bate10} for alternative  GPU microlensing algorithms and for more discussion about the dependence of the GPU performance on the parameter space explored.

Besides gravitational microlensing, there are other computational astrophysics calculations which the Xeon Phi coprocessor can significantly speed up.
One example is the Kerr black hole ray-tracing \citep{Schnittman04,Dexter2009,Kuchelmeister2012,Chan13,Chen13a,Chen15,Dexter16}.
Generating the image of an accretion disk strongly lensed by a central black hole using backward ray-tracing is an embarrassingly parallel computational problem, and it does not involve large data transfer or many memory operations.
Consequently, the Xeon Phi might perform very well.
Another good candidate is Monte Carlo radiation transfer \citep{Schnittman13a} which is also highly parallel in nature.
How much the performance of these simulations can be improved by using the Xeon Phi is a question for future work.
Given that future releases of the Xeon Phi have been already scheduled (e.g., Knights Hill and Knights Mill in 2017+) and will be deployed in several large national labs (for example, the KNC is already available at the ``Stampede Cluster" at the Texas Advanced Computing Center, the KNL is being tested and will be deployed for Phase II of the supercomputer ``Cori" at the NERSC, and the Knights Hill will be deployed for the supercomputer ``Aurora" at the Argonne National Lab), 
it is important for astronomers and astrophysicists to be aware of such computing possibilities offered by the Xeon Phi.
We conclude that the Xeon Phi is a very promising alternative to GPUs for high performance astronomical simulations.

\section{Acknowledgements}

The majority of the simulation was performed on the HPC cluster at the Research Computer Center at the Florida State Univ.
This research used resources of the National Energy Research Scientific Computing Center (NERSC), which is supported by the Office of Science of the U.S.  Department of Energy under Contract No.\  DE-AC02-05CH11231.
The authors acknowledge the Texas Advanced Computing Center (TACC) at The University of Texas at Austin for providing HPC resources that have contributed to the research results reported within this paper.
B.~C. thanks Prasad Maddumage and Edson Manners for discussions about Xeon Phi and GPU.
The authors thank the anonymous referee for a careful review of this work.

\bibliographystyle{model2-names}
\biboptions{authoryear}


\end{document}